\documentclass[conference]{IEEEtran}

\usepackage{graphicx}  

\begin{document}

\title{Strongly Coupled Plasmas in High-Energy Physics}

\author{\authorblockN{Markus H.Thoma}
\authorblockA{Centre for Interdisciplinary Plasma Science\\
Max-Planck-Institut f\"ur extraterrestrische Physik\\
P.O. Box 1312, D-85741 Garching, Germany\\
Email: thoma@mpe.mpg.de}}

\maketitle

\begin{abstract}
One of the main activities in high-energy and nuclear physics is the
search for the so-called quark-gluon plasma, a new state of matter
which should have existed a few microseconds after the Big Bang.
A quark-gluon plasma consists of free color charges, i.e. quarks and gluons,
interacting by the strong (instead of electromagnetic) force.
Theoretical considerations predict that the critical temperature
for the phase transition from nuclear matter to a quark-gluon plasma
is about 150 - 200 MeV. 
In the laboratory such a temperature can be reached in a so-called
relativistic heavy-ion collision in accelerator experiments.
Using the color charge instead of the electric charge,
the Coulomb coupling parameter of such a system is of the order 10 - 30.
Hence the quark-gluon plasma is a strongly coupled, relativistic plasma,
in which also quantum effects are important. 
In the present work the experimental and theoretical status 
of the quark-gluon plasma physics will be reviewed, 
emphasizing the similarities and 
differences with usual plasma physics. Furthermore, 
the mixed phase consisting of free quarks and gluons together with
hadrons (e.g. pions) will be discussed, 
which can be regarded as a complex plasma due to
the finite extent of the hadrons. 

\end{abstract}

\IEEEpeerreviewmaketitle

\section{Basic Facts of Quark, Gluons, and Quantum Chromodynamics}

One of the main tasks of modern nuclear and high-energy physics is the
quest of the quark-gluon plasma (QGP), a new state of matter, 
predicted to exist 
at high temperatures and densities. Quarks have been proposed in the middle
of the sixties as elementary substructure particles of the strongly
interacting particles, the so called hadrons, in order
to explain the large spectrum of hadrons discovered at this time
\cite{Gell-Mann}. Quarks are fermions, carrying the spin 1/2. Nowadays,
6 different quark species, characterized by the quantum number flavor
(up, down, strange, charm, bottom, top) are known. These quark flavors
cover a wide regime of masses, from about 5 MeV for the up-quark to 170 GeV
for the top-quark. In addition, quarks have an electric charge, either
2/3 or -1/3 of the electron charge. 

There are two different combinations of quarks for building hadrons:
baryons, i.e. hadrons which are fermions, e.g. protons
and neutrons, are fomed by 3 quarks, 
whereas mesons, which are bosons, e.g. pions, consist of a 
quark and an anti-quark. Atomic nuclei contain only protons and neutrons,
which are build up of up- and down-quarks. The other (heavier) quark flavors
exist only in short-living hadrons, created in accelerator experiments
or cosmic rays.
So far no free quarks have been observed in nature, a phenomenon called 
confinement. 

For explaining the hadron spectrum as well as cross sections
for the hadron production in electron-positron annihilations, additional
quark states had to be introduced. These states are characterized by the
quantum number color, which exists in three different possibilities
(red, blue, green). In the middle of the seventies, it was realized, that
this quantum number can be regarded as the charge of the strong interaction
between the quarks. This led to the formulation of a theory for the
quark interaction, called Quantum Chromodynamics (QCD) \cite{Leutwyler}, 
analogously
to QED, describing the electromagnetic interaction. For example, in QED
the scattering of two electrons is mediated by the exchange of a photon.
The exchange particles of QCD, mediating the strong interaction, are
called gluons. The main difference between QED and QCD is the fact that 
quarks come in three different colors, whereas electrons carry only 
one charge. Mathematically this implies that QED is based on the 
abelian $U(1)$ gauge group, while QCD on the non-abelian $SU(3)$.
As a consequence, gluons carry a color charge, in contrast to photons which
have no electric charge. Therefore, gluons can interact directly with
other gluons. A consequence of this gluon self-interaction is the
fact that, in contrast to QED, the interaction strength of two
color charges increases with the distance between these partons.
(A parton is either a quark or gluon.) Then, for instance, a meson can be
pictured as a rubber band with a quark on one end and an anti-quark on 
the other. If the quarks are close together, they do not feel each other.
This phenomenon is called asymptotic freedom, which can be derived
directly from QCD. If the distance between the quark and the anti-quark
gets larger, however, the force between them increases, until the rubber band
breaks into two new mesons, e.g. in 
high-energy scattering experiments. In this way, one can explain confinement: 
the interaction between hadrons can lead 
only to new hadrons but not to freely propagating quarks. In other words,
only colorless states, the hadrons, can exist as free particles. However,
it should be noted that confinement has not been proved rigorously from
first principles. 

Following this picture, hadrons can be considered as quark bags in which
the quarks are confined by their interaction with each other. Since the 
interaction is mediated by gluons which also can split into virtual
quark anti-quark pairs, the hadrons are complicated systems containing
quarks and gluons (see Fig.1). A nucleus is now represented by a dense system
of individual quark bags, the protons and neutrons. 

\section{Quark-Gluon Plasma: Experiment and Theory}

Shortly after quarks were proposed the question arose, what will happen
if we compress the nucleus, i.e. increase the nuclear or baryon density,
or if we heat up the nucleus in a fixed volume
to such high temperatures that pair creation 
(e.g. of pions) occurs in scattering events of hadrons
with large thermal energies. 
Then the system becomes denser and the bags should start
to overlap and to dissolve, transforming the nucleus into 
a ``soup'' of freely propagating quarks and gluons as sketched in Fig.2, 
which was called
a quark-gluon plasma \cite{Mueller}. 
Theoretical estimates
predicted a critical baryon density of about 10 times normal 
nuclear matter density, $\rho_0=0.125$ Gev/fm$^3$ corresponding
to $2.2 \times 10^{17}$ kg/m$^3$. The critical temperature $T_c$ 
of this deconfinement phase transition from hadronic matter to the QGP phase
should be of the order 150 to 200 MeV (1.8 to $2.4 \times 10^{12}$ K).

\begin{figure}
\centering
\includegraphics[width=2.0in]{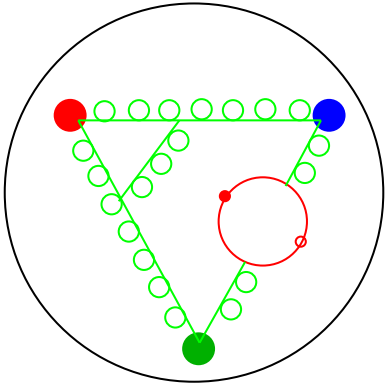}
\caption{A baryon as quark bag containing three quarks interacting via gluon
exchange}
\label{fig1}
\end{figure}

The phase diagram of nuclear matter is pictured in Fig.3, where the phase 
transition line in the plane given by the temperature and baryon density
is shown. Also three different possibilities to cross this border line
are indicated. Until about a few $\mu$s after the Big Bang the Early Universe
had a temperature above the critical temperature and should have existed 
in the QGP phase. In the center of neutron stars, on the other hand,
the baryon density might exceed the critical one at low temperatures. 
Then the interior
of these compact stars might actually consist of degenerate
quark matter instead of neutron matter\footnote{As a matter of fact, recent 
investigations revealed that there might be a quark pairing (color 
superconductivity) at high quark densities, which could have an influence on 
the properties of neutron stars containing quark matter (see e.g. 
Ref.\cite{Alford}).}. Finally in high-energy
nucleus-nucleus collisions, also called relativistic heavy-ion collisions,
a dense and hot fireball is created for a short time, which might be
also in the QGP phase \cite{QMproc}. 

\begin{figure}
\centering
\includegraphics[width=3.5in]{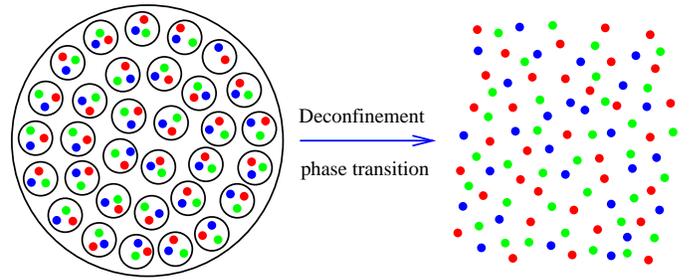}
\caption{The deconfinement phase transition from nuclear matter to QGP}
\label{fig2}
\end{figure}

Different accelerator experiments have been and will be performed for
searching for the QGP. 
For producing a large and hot fireball, heavy-ion 
collisions with Au-, Pb-, and U-beams at center of mass energies
from 4 to 5500 GeV per nucleon have been or will be conducted
at the Brookhaven National Laboratory on Long Island (AGS, RHIC) and
at CERN in Switzerland (SPS, LHC). At the accelerator SPS first
indications for the discovery of the QGP plasma have been found and 
announced at CERN in
2000 \cite{Heinz}. Also in 2000 the dedicated heavy-ion experiment RHIC
started its operation with energies up to 200 GeV per nucleon, which is about
10 times the energy of SPS. Due to the huge amount of data - thousands of
particle tracks are produced in a single collision - the analysis
of the data is still in progress, although preliminary results confirm
the conclusions from SPS. The high density phase at moderate temperatures
will be addressed at a future accelerator facility at the GSI Darmstadt, which 
eventually may allow to study the superconducting phase of QCD.

Theoretical estimates for these collisions predicted a maximum temperature
clearly above the critical at RHIC ($T_{max}> 230$ MeV), a maximum
volume of the hot fireball of about 3000 fm$^3$, containing about
10000 partons, and a life-time of the QGP of about 5-10 fm/c (1 fm/c
corresponds to $3\times 10^{-24}$ s). This short life-time is caused by the 
fast expansion of the fireball leading to a fast reduction of
the maximum temperature and the phase transition to the hadronic state.
The pre-equilibrium phase, i.e.,
the time needed to form an equilibrated fireball after the collision of
the cold nuclei, was estimated to be of the order of only 1 fm/c due to the
strong interaction between the particles. Hence, the chances to create
an equilibrated QGP in relativistic heavy-ion collisions seem to be
good. 

However, the main problem lies in the detection of this state
since the QGP, living only for an extremely short period in a small
spatial volume, cannot be observed directly. Only by comparing theoretical 
predictions for signatures of the QGP, such as the spectra of the emitted
particles (photons, leptons, and hadrons) which reach the detector
after the decay of the expanding fireball, with experimental data,
the QGP can be discovered. Unfortunately, so far 
no unambiguous signature has been proposed due to the complexity of
the strongly interacting many-body system. Therefore, the final identification
of the QGP might not be possible by a unique signature but rather by
circumstantial evidence owing to a combination of all possible signatures.

\begin{figure}
\centering
\includegraphics[width=3.0in]{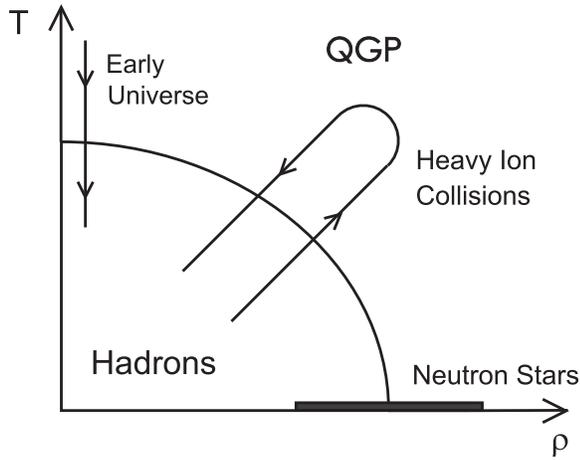}
\caption{Phase diagram of nuclear matter (temperature vs. baryon density)}
\label{fig3}
\end{figure}

The theoretical description of the QGP has to take into account that it is a 
strongly coupled system at temperatures reachable in heavy-ion collisions.
At $T\simeq 200$ MeV the strong coupling constant $g$, corresponding to the
electromagnetic coupling $e$, assumes values of $g\simeq 1.5$ - $2.5$ 
in natural
units ($\hbar =c=k_B=1$). The strong coupling constant is temperature dependent
and decreases with increasing temperature and density
because of asymptotic freedom, i.e. the QGP becomes 
an ideal plasma in the infinitely high temperature limit. 
At $T\simeq 200$ MeV, however, the QGP is a non-ideal plasma. Its Coulomb
coupling parameter is given by the chromoelectrostatic interaction energy
divided by the thermal parton energy\footnote{Of course, in the 
relativistic case, the (chromo)magnetic interaction becomes equally 
important.}

\begin{equation}
\Gamma=\frac{Cg^2}{\Delta T},
\label{eq1}
\end{equation}

where $\Delta $ is the distance between the partons, following approximately
from the
number density $n\sim T^3$ assuming the Stefan-Boltzmann limit
(ideal gas of massless particles). At $T=200$ MeV the distance is of the order
$\Delta = 0.5$ fm. The constant $C=3$ for gluons and 4/3 for quarks come from
the different color charges of gluons and quarks, respectively. Typical values
of $\Gamma$ for temperatures, which can be achieved in heavy-ion collisions, 
are between 10 and 30, corresponding to a strongly coupled plasma. For 
$T\rightarrow \infty$ the coupling parameter approaches zero, although
$\Delta \sim 1/T$ since $g(T=\infty)=0$, which demonstrates the ideal
plasma limit at high temperatures. 

Furthermore, the QGP is a relativistic system as the masses of its components
are much smaller than their thermal energies which are
of the order of the temperature. Gluons are massless, while the up- and
down-quarks have masses of a few MeV. Only the mass of the strange-quark 
($m_s\simeq 150$ MeV) is comparable to the temperature. The other
quark flavors have much higher masses and are therefore not thermally
exited, i.e., they are suppressed compared to up-, down, and strange-quarks 
in the QGP. 

Finally, at temperatures
not too far from the critical, quantum effects play an
important role in the QGP. Hence a theoretical description has to start
from a statistical quantum field theory, i.e. thermal QCD. Due to the large
coupling constant non-perturbative methods are required. The only 
well-established non-perturbative approach for quantum field theories are
lattice simulations, where the quantum theoretical equations of motions
are solved numerically on a 4 dimensional space-time grid using
Monte-Carlo techniques \cite{Karsch}. 
This method has the advantage that it works for all
temperatures below as well as above the phase transition, allowing, for 
example,
to determine the critical temperature, the order of the phase transition,
or the equation of state of the QGP. However, it is very difficult to compute
dynamical quantities such as most of the signatures of the QGP which are based
on particle production rates or similar dynamical processes. Hence,
perturbative methods, extended to finite temperatures and 
improved by resummation techniques, are also adopted 
to estimate dynamical properties and signatures of the QGP \cite{Thoma}.
Although they might be valid only at temperatures far above the critical one, 
these methods give at least important insights in the dynamical structure
of the QGP.

In the high temperature limit ($T\gg T_c$ corresponding to $g \ll 1$), 
perturbation theory can be shown to be
equivalent to the semiclassical Vlasov equation \cite{Elze}, generalized
to the relativistic case and the strong instead of electromagnetic
interaction. The only quantum input comes from using the Bose distribution
for gluons and the Fermi distribution for quarks instead of the Maxwell
distribution. As in usual plasma physics one can derive, for instance,
the dielectric tensor from the Vlasov equation. In the case of an isotropic
and homogeneous system this tensor has only two independent
components (longitudinal and transverse dielectric functions), which depend
on the energy (frequency) $\omega$ and momentum (wave number) $k$:

\begin{eqnarray}
\epsilon_L(\omega, k) & = & 
1+\frac{3m_g^2}{k^2}\> 
\left (1-\frac{\omega}{2k}\, \ln \frac{\omega+k}{\omega-k}\right ),
\label{eq2}\\
\epsilon_T(\omega ,k) 
&=& 1-\frac{3m_g^2}{k^2}\> 
\left [1-\left (1-\frac{k^2}{\omega^2}\right )\, 
\frac{\omega}{2k}\, \ln \frac{\omega+k}{\omega-k}\right ],\nonumber
\end{eqnarray}

where $m_g^2=g^2T^2/3\, (1+N_f/6)$ is called the effective gluon mass
with $N_f$ thermally excited quark flavors in the plasma. The same result 
can be obtained from the gluon self-energy in
lowest order perturbation theory in the high temperature
limit. Going beyond this approximation, perturbative QCD allows a systematic
treatment of higher order effects, such as damping rates coming from
collisions between the particles. However, in order to avoid infrared
singularities due to the exchange of massless gluons and violations
of the gauge invariance, resummation techniques have to be adopted at finite
temperature \cite{Thoma}. The development of such methods is a very
active research topic. 

\begin{figure}
\centering
\includegraphics[width=4.0in]{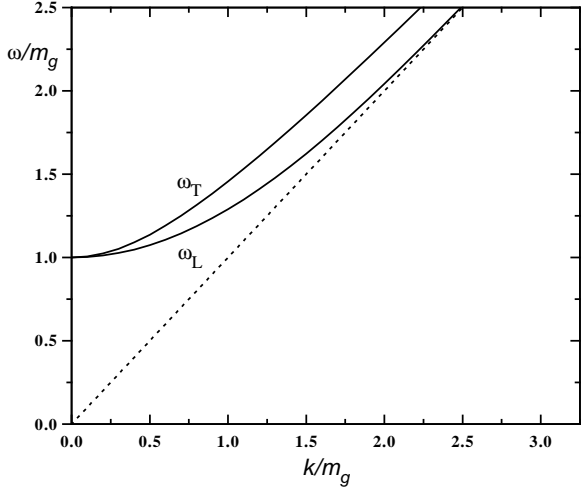}
\caption{Dispersion relation of longitudinal and transverse plasma waves in 
the QGP}
\label{fig4}
\end{figure}

From the chromoelectromagnetic Maxwell equations one obtains the constraints
on the dielectric functions,   
$\epsilon_L(\omega ,k)=0$ and $\epsilon_T(\omega ,k)=k^2/\omega^2$,
which describe the propagation of collective plasma modes. From these
equations dispersion relations $\omega_{L,T} (k)$ for longitudinal and 
transverse plasma modes
in the QGP follow as shown in Fig.4. Both branches start at $k=0$ with
the same energy $\omega_{L,T}(k=0)=m_g$, which can be identified with
the plasma frequency. The longitudinal branch is known as plasmon. In the 
case of relativistic plasmas both branches are equally important, whereas
in non-relativistic plasmas the transverse mode is usually neglected.
Both modes approach the vacuum dispersion $\omega =k$ for large momenta. 
However, the spectral strength of the plasmon mode vanishes exponentially
for large momenta, $k\gg gT$. 
Debye screening follows from the static limit ($\omega =0$)
of the longitudinal dielectric function describing the modification
of the Coulomb potential to a Yukawa potential
in the presence of a medium. The Debye screening
length is given by $\lambda_D=1/(\sqrt{3}m_g)$. The dielectric functions
of (\ref{eq2}) also contain Landau damping for $\omega ^2 <k^2$, for
which the logarithms become imaginary. Although the plasma waves
in the semiclassical approximation are located in the region $\omega^2>k^2$
(see Fig.4), where no Landau damping takes place, Landau damping of 
virtual gluons (plasma waves) in the QGP with $\omega ^2<k^2$ 
plays an important role in higher order processes.

A further very interesting property of ultrarelativistic plasmas is
the existence of collective fermion modes in analogy to the plasma modes, i.e.
collective quark modes in the QGP.
They can be derived within perturbation theory from the in-medium quark
self-energy \cite{Peshier}.

Finally, it is interesting to note that in the high temperature limit 
the non-abelian effects are negligible. Hence, apart from numerical factors, 
the above results hold also for an ultrarelativistic QED plasma, i.e. an 
electron-positron plasma in which the temperature is much higher than the
electron rest mass. Such a situation is realized in Supernovae explosions, 
where temperatures up to 30 MeV are possible. The dielectric functions
of such a system are given by the above equation (\ref{eq2}), in which
one only has to replace the effective gluon mass $m_g$ by an effective
photon mass $m_\gamma = eT/3$.

\section{Thermophoretic Flow in the Mixed Phase} 

As a last point we want to consider a possible mixed phase of a QGP and
hadrons. Such a mixed phase would occur if the phase transition from
the QGP to the hadronic phase were of first order. During the expansion of the 
fireball in a heavy-ion collision or in the Early Universe, the mixed phase
is reached, once the temperature has 
dropped to the critical. Then bubbles of
hadrons will form (hadronization) 
and grow in the QGP until the entire phase consists of 
hadrons. QCD lattice simulations show that the
order of the phase transition depends sensitively on the values of the
up-, down-, and strange-quark 
masses. The latest calculations show that the QCD phase 
transition is either of first order or continuous, with a preference for the 
latter case \cite{Karsch}.

\begin{figure}
\centering
\includegraphics[width=2.5in]{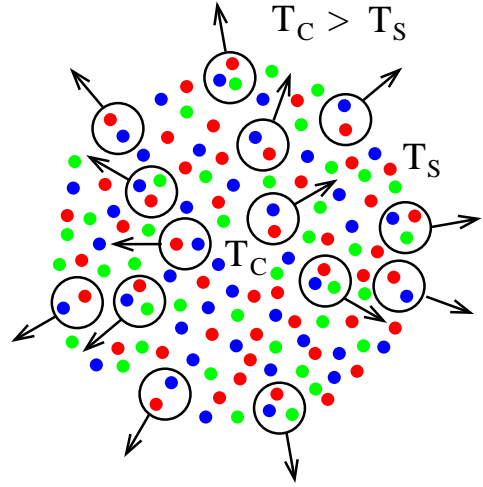}
\caption{Thermophoretic flow of hadrons in the mixed phase of the fireball}
\label{fig5}
\end{figure}

Assuming a first order phase transition, the mixed phase can be considered
as a complex plasma, in which extended and composite objects, the hadrons, are
embedded in the QGP containing point-like partons, similar as dust grains in
a low-temperature plasma. The main difference lies in the fact that the dust
particles are highly charged, whereas the hadrons are color-neutral
objects. Now, we assume in addition to the mixed phase a temperature
gradient in the fireball from the center to the surface and, for simplicity, 
a radial expansion of the fireball. 
Then, similar as in complex plasmas, a thermophoretic force will act
on the hadrons, pushing them from the center to the surface as shown in Fig.5.
The thermophoretic force in a complex plasma has been estimated from simple
kinetic arguments and shown to be in excellent agreement with experimental
results \cite{Rothermel}. We have generalized the kinetic derivation
of the thermophoretic force to the relativistic case and calculated
the final velocity of the hadrons from this effect. Even for small
temperature gradients of a few MeV/fm and small life-times of the
mixed phase (1-2 fm/c), we obtained large outward flow velocities
of the order of 0.8 c for pions and 0.4 c for protons \cite{Thoma1}. These
values are larger than the typical flow velocities of 0.5 c,
observed in relativistic heavy-ion collisions at SPS. 
We take this as an indication for the absence of a mixed phase in these
experiments, either due to a continuous phase transition instead of
first order or due to an absence of the QGP phase at all.

%\begin{figure}
%\centering
%\includegraphics[width=2.5in]{myfigure}
% where an .eps filename suffix will be assumed under latex, 
% and a .pdf suffix will be assumed for pdflatex
%\caption{Simulation Results}
%\label{fig_sim}
%\end{figure}

\end{document}